# INTERACTION BETWEEN ATOMIC PROJECTILES AND A CRYSTAL SURFACE AT GRAZING INCIDENCE: A COMPUTER SIMULATION

V.S. Malyshevsky*, G.V. Fomin

Southern Federal University, Rostov-on-Don, 344090 Russia

E-mail: vsmalyshevsky@sfedu.ru

**Abstract**

Features of the angular distributions of accelerated atomic projectiles at grazing angles of incidence on the crystal surface are studied by using the computer simulation. The interaction between the projectiles and the crystal-lattice atoms and atomic structure of the crystal surface are calculated by means of the electron density functional method. The angular distributions of scattered projectiles are simulated by taking into account their interaction with several atomic layers in the crystal lattice and atomic thermal displacements. Good agreement between the calculated results and known experimental data is achieved. A possibility of reconstructing the ion-atom dynamic interaction potential from the dependence of the rainbow scattering angle of nitrogen atoms under conditions of grazing incidence on the surface of an aluminum crystal on the total kinetic energy of the accelerated atomic particles is shown.

1. ## Introduction

In the past decades the grazing angle incidence geometry has become popular to probe surface properties (see, for instance, [1,2]). The so-called rainbow scattering in the case of the grazing incidence of particles on the surface of crystals has been the subject of numerous studies [3–9] for many years. This phenomenon occurs in the direction of the particle motion at a small angle along atomic chains, when the dependence of the azimuthal scattering angle (i.e. in the plane perpendicular to the atomic chains) on the impact parameter relative to the chosen atomic chain has extrema. The scattering angle, which corresponds to the maximum angular distribution of scattered particles, is termed 'the rainbow angle'. This effect is curious from the view point that the angular distribution of scattered particles provides information on the surface structure and the interaction potential between scattered atoms and the crystal surface. It is worth mentioning that rainbow scattering is also experimentally confirmed for particles passing through crystals[10]. Similarly with the grazing incidence on a crystal surface, the rainbow effect is observed due to the nonmonotonic dependence of the scattering angle on the impact parameter.

Some features have been discovered in the scattering of neutral atoms with energies up to several dozen kiloelectronvolts at the grazing incidence on crystal surfaces. One of these features reported in work [3], consists in a differences in the orientation dependence of the scattering of neutral atoms of some inert gas atoms and metals, which is due to the polarization of the incident neutral projectiles [11]. A quantitative explanation of this phenomenon depends on a correct selection of the far-range part of the interaction potential. [12] Another feature [4] consists in the nontrivial dependence of the rainbow scattering angle on the total energy of atoms that incident on the surface of metal and dielectric crystals. According to this, at a fixed initial transverse energy value, the angle of rainbow scattering from a metal surface decreases with increasing total initial energy. This effect is only observed for the scattering from metal surfaces and is absent in the case of scattering from dielectrics.

## 2. Interaction Potential

To construct the pair interaction potential between the incident projectiles and crystal surface atoms and also to study the structure of the single crystal surface layer, we carried out a quantum mechanical calculation of the total energy of the system's ground state, which includes the energy of electron–electron and electron–nucleus interactions for fixed positions of nuclei. The dependence of the total system energy on the interatomic distance corresponds to the sought interaction potential in the adiabatic approximation for an equilibrium electron subsystem. We define the potential of the interaction between an incident atom and an oriented crystal surface as a superposition of the potentials of the interaction between individual surface atoms (ions) in the first and subsequent layers

$$U(\mathbf{r}) = \sum_j V_a(|\mathbf{r} - \mathbf{r_j}|). \tag{1}$$

Quantum mechanical calculations of the pair potential are performed by using the theory of the electron density functional [13, 14], which recently underwent a rapid development. The main idea of this method involves a passage from the many-electron wave function of a system of electrons to the electron density function, which, together with developed exchange–correlation functional, enables us to describe the energy and geometry of the ground state of multi-particle atomic systems with high accuracy and without using empirical information. In the present paper, we used the so-called approximation of the local electron density [15, 16], which enables us to calculate the binding energy with an accuracy of ~10% and the interatomic distances with an accuracy of ~1%. Calculations were carried out using the *abinit* program package [17–20], in which Kohn–Sham wave functions were expanded in terms of the basis of plane waves and electrons of atomic cores were taken into account using norm conserving pseudopotentials calculated in terms of the Troullier–Martins scheme [21]. In particular, the electron density distribution of two interacting atoms Al and Na at distances of 2 Å and 4 Å, respectively, is shown in Figure 1.

Comparison of the pair interaction potential of Al atoms with the classical Thomas–Fermi potential for different parameterization shows good agreement for small and large distances. In the range of impact parameters corresponding to the Al–Al binding length, there is a local potential minimum, which is due to the nonzero polarizability of neutral Al atoms, i.e., the long range attraction component of the interaction potential (Figure 2). For purposes of computer simulation of the trajectories of the atomic projectiles scattered at the crystal surface, it is convenient to use the well-known analytical three-parameter approximation (so-called Morse potential [22]):

$$V_a(r) = \varepsilon \left\{ \exp[-2\alpha(r-\sigma)] - 2\exp[-\alpha(r-\sigma)] \right\}, \tag{2}$$

that well describes the potential calculated by using electron density functional theory in the range of impact parameters being of interest to us in the case of the corresponding choice of the fitting parameters $\varepsilon$, $\alpha$, and $\sigma$. In Figure 1, the calculated Al–Al pair potentials are compared with the Morse and Moliere potentials, which are usually used in problems of atom scattering. In our further calculations, we shall use the parameters of the analytical Morse potential for the Al–Al as shown in Table 1.

The calculated structure of a cluster consisting of Al atoms near the single crystal surface layer was shown in work [8]. The error in determining the Al lattice constant was at most ~0.001 Å. As follows from the calculations [8], the distance between the first and second layers of the Al single crystal (001) plane was 1.743 Å as a result of a normal relaxation, while the distance between the nearest layers in the volume was 2.025 Å. Taking such an approach to the second layer surface into account can significantly affect the angular distribution of particles scattered at the surface.

### 3. Simulation Procedure

Beam scattering was simulated using the Dormand-Prince method [23] of the numerical integration of classical non-relativistic equations of the movement for each individual particle in the field of an aluminum crystal, taking the two-particle interaction potentials into account. The field was assumed to be induced by the closest surrounding at selected numbers of atoms and layers of a crystal surface. The atomic positions of the crystal corresponded to the known crystal structure and dispersion of the normal heat-displacement distribution. Each individual particle in the beam over the whole path length on the order of several lattice constants was impacted by a field of static atoms at randomly displaced positions.

The interaction potential of the incident atom with the oriented crystal surface was defined as the superposition of the potentials of the interaction with individual atoms (ions) of the surface in the first and subsequent layers. The interaction of the atomic projectiles with the crystal atoms was described by the Morse potential. As is shown above, the Morse approximation in Equation 2 is suitable to describe the static potential calculated in the context of density functional theory over a wide range of impact parameters.

The angular distribution of rainbow scattering depends on the structural features related to the relaxation and reconstruction of the surface. The normal relaxation arises in metals, and there is frequently a decrease in the first interlayer distance. This circumstance causes modifications in the potential relief scattering of the particles incident on the surface, as well as changes in their trajectories and angular distributions. These and other peculiarities were taken into account in the construction of the model of the interaction and scattering of atomic particles at the crystal surface [8]. The output parameters of the scattered particles were fixed using polar coordinates in accordance with the technique reported in work [5]. The developed application enables us to track the individual trajectories of the scattered particles and the output parameters as a function of the impact parameter. To obtain enough statistics, data were collected for more than 20000 trajectories.

The rainbow scattering phenomenon is attributed to the case in which the dependence of the scattering angle on the impact parameter is nonmonotonic. In the case of particle scattering by the crystal surface during the orientation motion along atomic chains (the scattering scheme is shown in Figure 3), the effect of rainbow scattering occurs when the dependence of the azimuthal scattering angle on the impact parameter with respect to a chosen atomic chain has extrema.

4. Polarization Effects

The results of measurements performed, for example, in work [3, 4] showed a considerable difference between the orientation dependences of the angular distributions of scattered atoms of noble gases and some metals. The reason for this is likely the polarization of neutral atoms incident on the target surface, which may lead not only to peculiarities of scattering, but also to new, previously unstudied effects. Indeed, long-range attractive forces are usually neglected when the studying of processes that require small impact parameters, in the case of which a considerable mutual penetration of electron shells of colliding particles occurs. To describe collisions with large impact parameters properly, the long-range part of the potential is to be taken into account; in particular, this part depends on the polarizability of the colliding particles. Precisely such a case occurs at the grazing incidence of accelerated neutral atoms on the crystal surface. For small angles of arrival at the crystal, large impact parameters begin to play an important role in the formation of the angular distribution of scattered particles. Because the polarizability of metal atoms considerably exceeds that of atoms of inertia gases, a significant polarization effect for metal atoms can naturally be expected (for example, the polarizability of He and Ar atoms are 1.39 and 11.1 au, respectively, and those of Na and K atoms are 157 and 250 a.u., respectively).

For a for a fixed transverse energy value the dependence of the azimuthally scattering angle on the impact parameter with respect to the chosen atomic chain has pronounced maxima (Figure 4), and this dependence demonstrates the effect of rainbow scattering by the system of atomic chains. Figure 4 shows the angular distribution (in the transverse plane) of Na, Al and Ar atoms scattered at the Al crystal (001) surface in the case of the 10 keV particle incidence along the axial ⟨011⟩ direction with a transverse energy of $\varepsilon$ = 3 eV. The effect of rainbow scattering manifested in the appearance of two side symmetrical peaks is clearly seen. It should be noted that in Figure 5, azimuthal and polar angles for convenience are shown at different scales. Therefore, the rainbow angle value is displayed incorrectly.

The dependence of the rainbow angle on the initial transverse energy for Al, Na, and Ar atoms turns out to be significantly different (Figure 6). In our calculations, we use the parameters of the analytical Morse potential for the Al–Al and Na-Al as those in Table 1. The rainbow angle begins to increase at small transverse energies of Al and Na atoms. Such a behavior can be explained by the fact that, as the initial transverse energy of incident particles decreases, the angular distribution forms by the scattering processes with large impact parameters, and the polarization contribution to the interaction potential turns out to be considerable. Because the polarizability of the Al and Na atoms is much larger than that of the Ar atoms, this effect is hardly manifested for the Ar atoms. Thus, a simulation of the scattering process performed by using the Thomas–Fermi potential in an O'Connor–Biersack approximation [24] provides the best agreement with the data measured for the Ar atoms.

Attention should be paid to another interesting effect that polarization effects lead to. Because of the effect of the incident atom polarization, the potential of the interaction between them and the surface has a minimum, and the incident atom losing the energy during collision with the crystal surface can be captured by the surface bound state in the potential well [25]. The atom in this state moves along the plane, slightly oscillating in the transverse direction (Figure 7a). We note that a similar phenomenon was considered in work [26], where this effect was considered due to the polarization of the target surface by a flying point-like charged particle. Computer simulation [25] shows that there are also particles trapped in the bulk of the crystal (Figure 7b). The probability of the capture in such a state depends on the initial impact parameter. Obviously, such trajectories do not contribute to the angular distribution of the particle reflected from the crystal surface.

## 5. Dynamic Effects

As is shown in experiments [3, 4] there is an interesting dependence of rainbow scattering on energy in a case of dielectric and metallic crystals. Accordingly to works [4, 5], the rainbow scattering angle for N and O atoms depends on the normal velocity component (i.e. on the so-called transverse energy) differently at high and low energies. There is also a difference in the dependences of the rainbow scattering angle on the total energy under the same conditions for the fixed transverse energy in metallic Al and dielectric LiF. The rainbow scattering angle decreases when the total energy in Al increases, while this energy remains almost constant in LiF [4]. The noted dependence, that is observed in Al, may be associated with an interaction between incident atoms and electronic gas of metal. A movement of atomic particles is influenced by an interaction between them and electronic gas because of its Fermi statistics. A velocity of atoms used in experiments ($\sim 4 \cdot 10^7$ cm/s) is much smaller than the Fermi velocity of electrons ($2 \cdot 10^8$ cm/s) in Al. Only a small amount of electrons near the Fermi surface is capable to change their state on interacting to slow moving atoms, and, consequently, decelerates these atoms. Increasing the atom energy leads to an increase of the number of the electrons near the Fermi surface and to the deceleration effect [27, 28]. This is the main reason for the rainbow-angle narrowing as it was shown in works [6, 7].

Another method for the low energetic atomic collisions description uses pair potentials that depend on the velocity [29]. These potentials take into account a kinetic energy transfer of atoms to electrons and enter the Thomas–Fermi–Dirac equations. Again, Fermi statistics manages electron states that take part in the energy transfer. Both explications consider the deceleration effect as an atom-energy transfer to electrons having states in the atom velocity-dependent range. And in that way they are similar, since both of them refer to the energy transfer to the electronic subsystem, and the domain of the allowed initial energy states is determined by the relative motion velocity of the colliding atomic particles.

The present results on the particle paths simulation show some features of the N neutral atoms reflection on the (001) and (111) surfaces of Al crystal with incidence in directions <100> and $<1\bar{1}0>$, respectively. The curves of dependence of the rainbow scattering angle on the initial transverse energy for nitrogen atoms are radically different as a function of the kinetic energy of the atoms. The rainbow angle decreases with increasing energy. The parameters of the pair potential approximation were induced by comparison with experimental data. The Morse potential dynamic parameters listed in Table 2 and showed in Figure 8 are fitted the experimental data the best way. The potential parameters from Eq. (2) depend on the kinetic energy. This dependence may be presented by approximate formulas:

$$\varepsilon = \varepsilon_1 + \varepsilon_2 \exp(-E/\varepsilon_3), \quad \sigma = \sigma_1 + \sigma_2 \exp(-E/\sigma_3) \ . \tag{3}$$

The values of the constants from Eqs. (3) are given in Table 3.

Two angle distributions are in good agreement with experimental results [4,5]. These distributions calculated by using parameters (2) with the transverse energy that equals 12 eV are shown in Figure 9. The rainbow scattering angle as a function of the transverse energy in different crystallographic orientations (Figure 10 and 11) is also consistent with experimental data [4,5].

However in the entire range of transverse energies, the Morse approximation is not valid for the N atom energy of 70 keV. At this energy, the atom velocity is approximately equal to the Fermi velocity of Al electronic gas. Consequently, all the existing electronic states taking part in the energy transfer have exhausted, and a further increase of a braking force with increasing the atom velocity is stopped. Starting from this velocity, the energy loss decreases with increasing the particle velocity due to the scatter short-living potential. Using the Morse potential approximation does not take into account the role of the atomic potentials screening at the low transverse energies and order velocities being the Fermi velocity order. Obviously, a lower scattering angle supposes a smaller force acting on the particle or a larger screening of atomic potentials. The scattering process is now described by a repulsive potential determined from the interaction between the nuclei and atom inner-shell electrons. The simulation for this process is performed by using the O'Connor–Biersack approximation [24] and leads to good agreement with the data measured for the atomic energy (70 keV) in the entire transverse-energy range.

Thus, the result of our study [30] shows that the Morse dynamic potential approximation is effective in describing the atom particles rainbow scattering for a special range of parameters. Namely, we have determined the parameters of the pair interaction potential that are the best-way fitted with experimental data for the energy range between 10 and 70 keV of N atoms incident on the Al crystal surface.

## 6. Conclusions

In conclusion, we have presented an analysis of rainbow structures observed in the scattering of fast atoms from oriented crystal surfaces. Our computer simulation shows that polarization effects can play an important role in describing the processes of the reflection of neutral atoms from the surface. Because the atomic polarizability of metals greatly exceeds that of inert gases, a significant polarization effect can be observed for precisely them. Ability to describe the rainbow scattering of atomic particles from the surface of metal in the case of the grazing incidence by means of the dynamic potential in the Morse approximation has been shown in the present study. The pair-potential interaction parameters for accelerated neutral nitrogen atoms have been calculated via simulating the dependence of the rainbow scattering angle on the energy of aluminum particles incident on the crystal surface. Our simulation results have been found to be consistent with known experimental data [3-5, 12]. A more thorough analysis of the energy dependence of the rainbow scattering of atomic particles from crystal surfaces will be possible at the implementation of the idea of the experimental separation of the nuclear and electron losses. This task has not yet been solved by measuring retardation losses in transmission experiments.

**Acknowledgements**

This work is supported by the Southern Federal University (Rostov-on-Don, Russia). Assistance by Prof. V.Yu. Topolov on improving the manuscript text is gratefully acknowledged. We enjoyed fruitful collaborations with Dr. L.A. Avakyan, T.I. Zhilina and E.V.Dergacheva.

TABLES

Table 1. Values of the Morse potential parameters for Al–Al and Na-Al

|  | Al | Na |
|---|---|---|
| $\varepsilon$, eV | 0.27 | 0.11 |
| $\sigma$, Å | 3.25 | 3.76 |
| $\alpha$, 1/Å | 1.16 | 0.97 |

Table 2. Values of the Morse potential parameters for N–Al at different Nitrogen atom energies

| $E$, keV | 10 | 20 | 25 | 35 | 40 | 70 |
|---|---|---|---|---|---|---|
| $\varepsilon$, eV | 0.41 | 0.37 | 0.35 | 0.32 | 0.30 | 0.25 |
| $\sigma$, Å | 2.76 | 2.88 | 2.95 | 3.05 | 3.10 | 3.30 |
| $\alpha$, 1/Å | 1.0 | 1.0 | 1.0 | 1.0 | 1.0 | 1.0 |

Table 3. Constants of the potential approximation

| $i$ | 1 | 2 | 3 |
|---|---|---|---|
| $\varepsilon_i$ | 0.189 | 0.276 | 45.74 |
| $\sigma_i$ | 3.616 | −1.016 | 59.74 |

FIGURE CAPTIONS

**Figure 1**. Electron density distribution of two interacting atoms Al and Na at distances of 2 Å (a) and 4 Å (b).

**Figure 2**. Calculated pair potentials of Al–Al interaction for different approximations.

**Figure 3**. Geometry of atom scattering at the crystal surface for grazing incidence.

**Figure 4**. Dependence of the azimuthally scattering angle on the impact parameter d (where $d_0$ is the distance between the nearest atomic chains) at incidence of 10 keV of Na and Ar neutral atoms on the Al surface along the ⟨011⟩ direction. The initial transverse energy is 3 eV. Symbols ■ and ○ indicate the scattering on single and double super-surface layers, respectively.

**Figure 5**. Angular distributions (in the transverse plane) of Na, Al and Ar atoms scattered at the Al crystal (001) surface in the case of 10 keV particle incidence along the axial ⟨011⟩ direction with a transverse energy of $\varepsilon = 3$ eV.

**Figure 6**. Dependence of the rainbow scattering angle of the neutral Na, Al and Ar atoms on the transverse energy during the incoming particle incidence with the energy 10 keV onto the (001) surface of an Al crystal along the <011> axial direction. Colored symbols ■, ● and ▲ indicate the experimental data [3, 4]. The respective simulation results for the same energy values are shown by the dash-dot lines.

**Figure 7**. Possible trajectories (a, c) of the neutral nitrogen atoms at grazing incidence onto the (001) surface of an Al crystal along the <100> axial direction and corresponding Poincaré sections (b, d).

**Figure 8**. Dynamic pair potential N–Al within the Morse approximation at different energies.

**Figure 9**. Results of computer simulations of the angular distribution of the scattered nitrogen atoms with the energies 10 and 70 keV in the polar coordinates under the conditions of grazing incidence of incoming particles onto the (001) surface of an Al crystal along the <100> axial direction. The initial transverse energy is 12 eV.

**Figure 10**. Dependence of the rainbow scattering angle of the neutral nitrogen atoms on the transverse energy during the incoming particle incidence with the energies 10, 25, 40 and 70 keV onto the (001) surface of an Al crystal along the <100> axial direction. Colored symbols ■, ●, ♦ and ▲ indicate the experimental data [4]. The respective simulation results for the same energy values are shown by the dash-dot lines. Dotted line – calculated rainbow angle at the energy 70 keV in the O'Connor–Biersack pair potential.

**Figure 11**. Dependence of the rainbow scattering angle of the neutral nitrogen atoms on the transverse energy during the incoming particle incidence with the energies 10, 20, and 35 keV onto the (111) surface of an Al crystal along the $<1\bar{1}0>$ axial direction. Colored symbols ●, ▲ and ■ are the experimental data [5]. The respective simulation results for the same energy values are shown by the dash-dot lines.

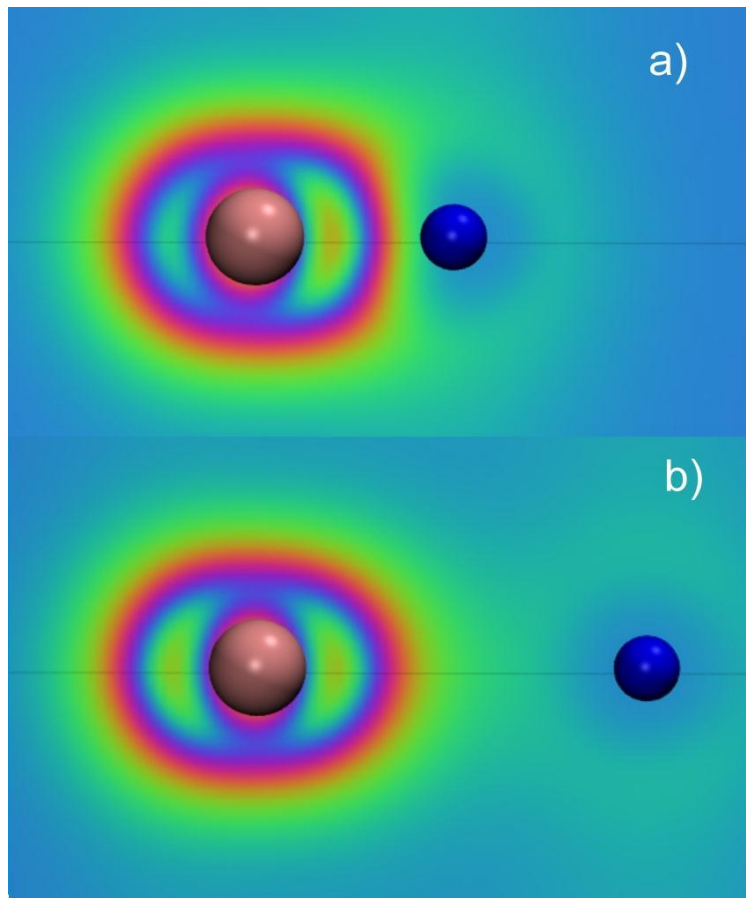

Figure 1

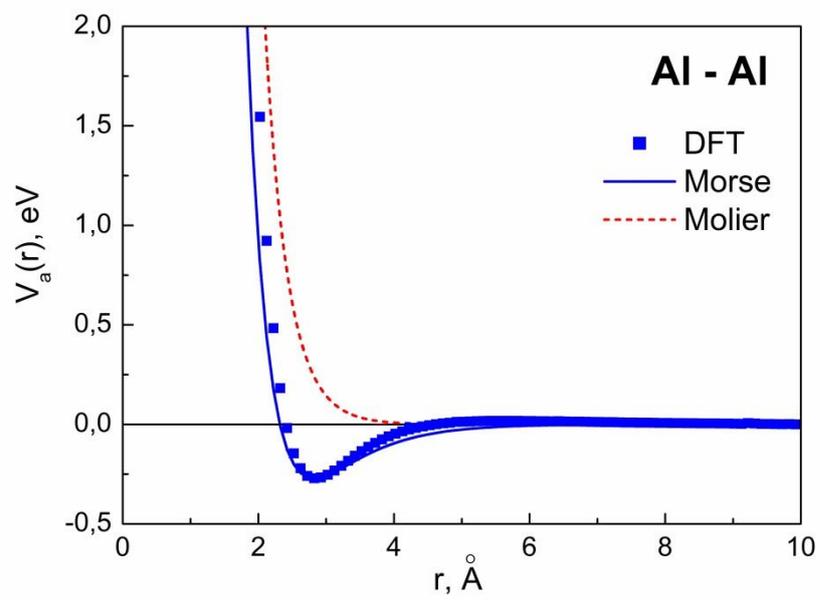

Figure 2

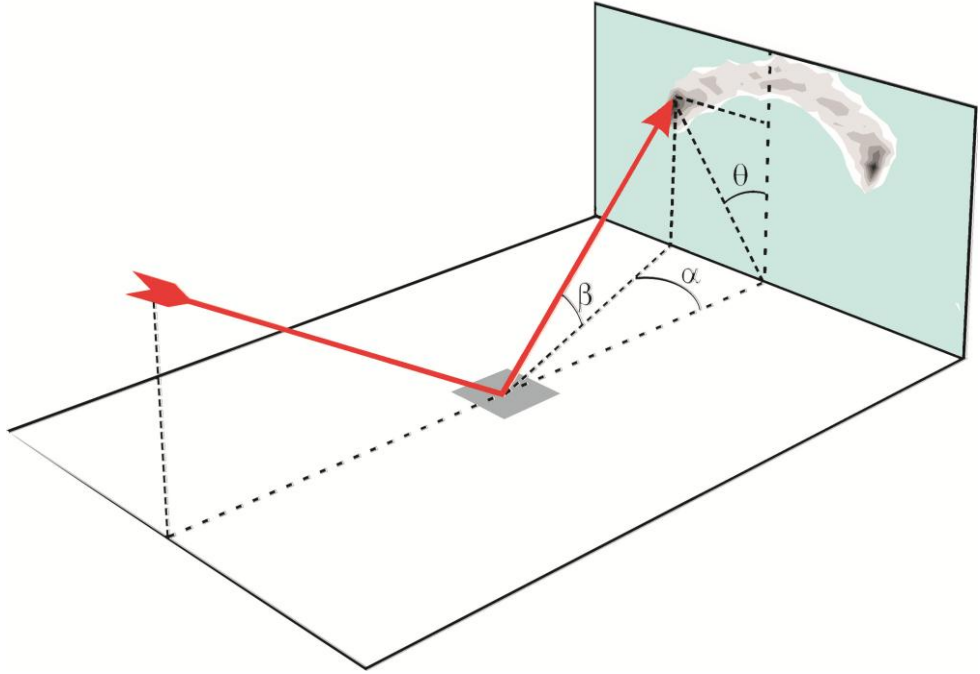

Figure 3

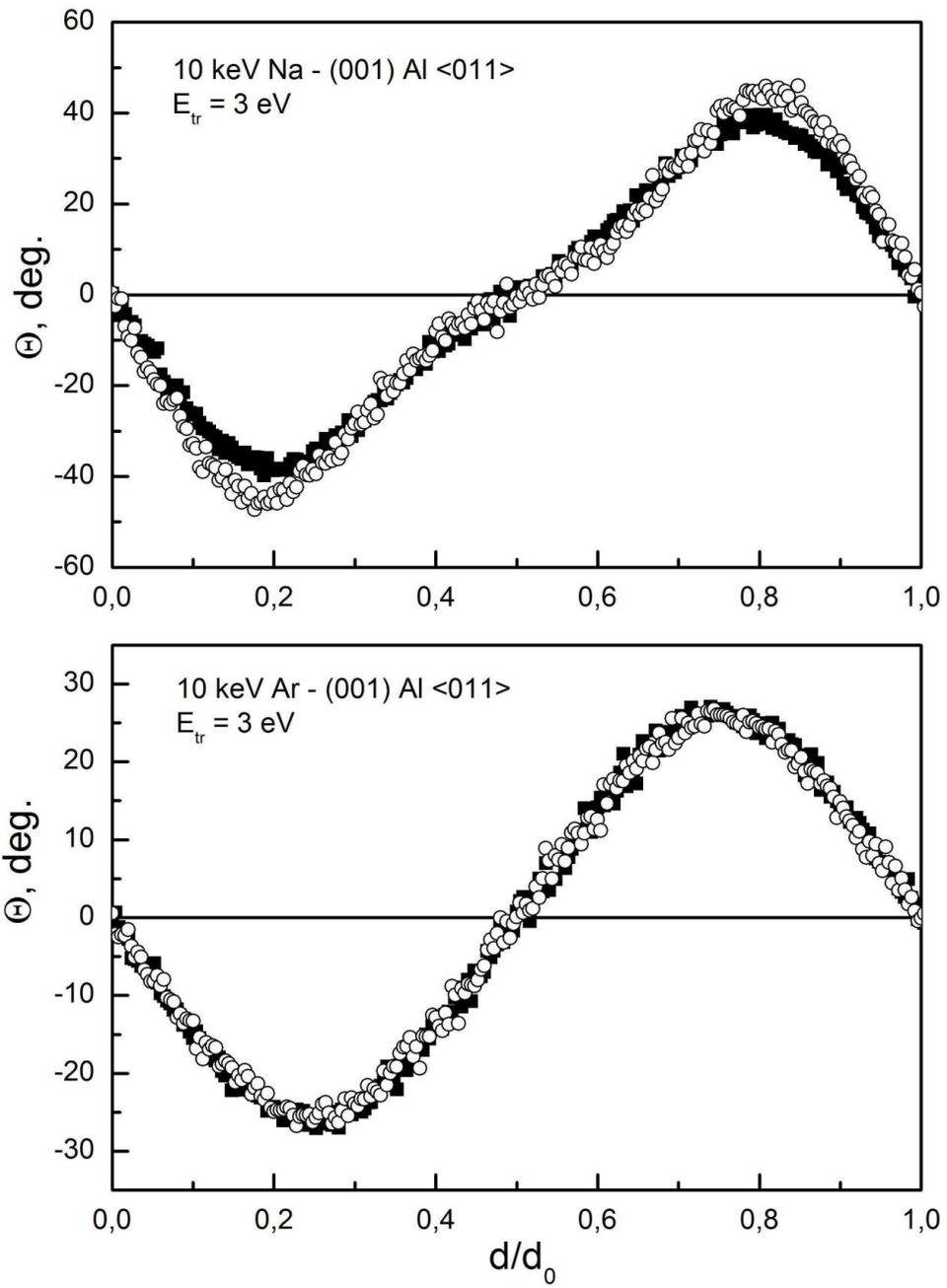

Figure 4

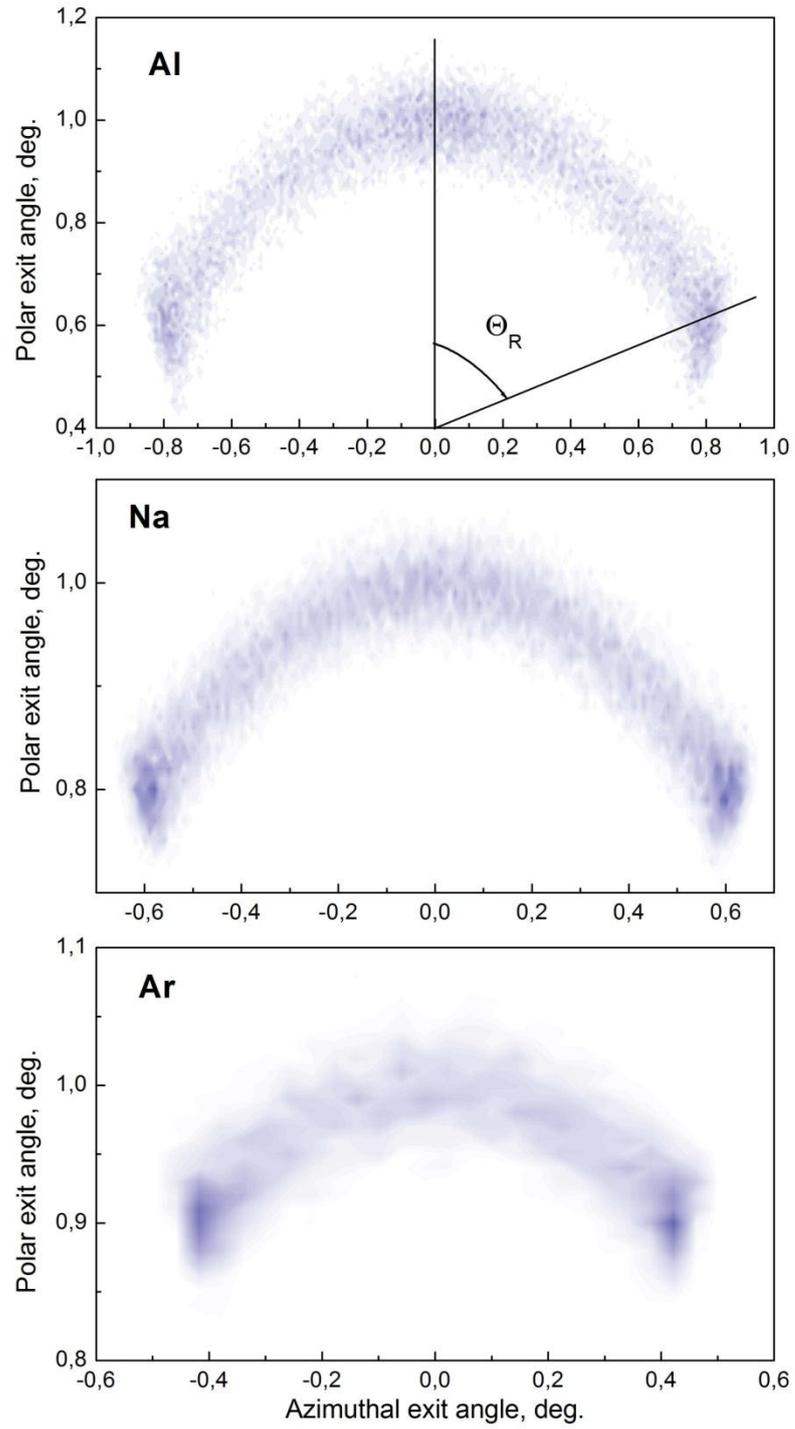

Figure 5

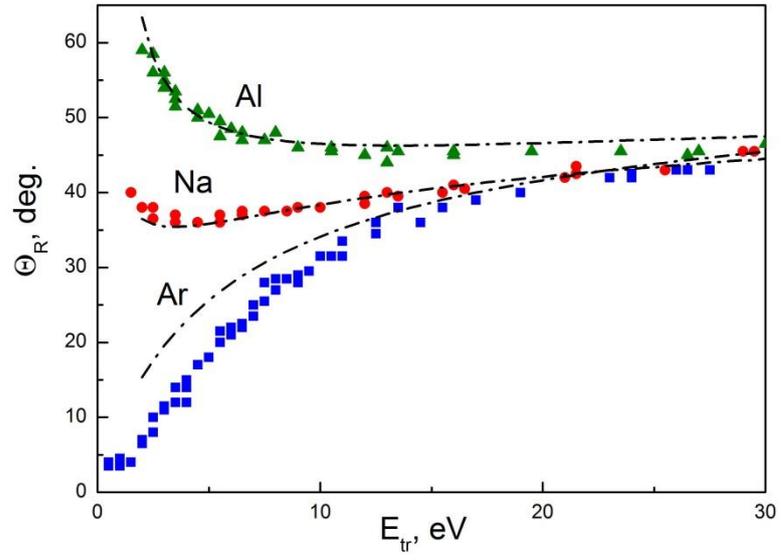

Figure 6

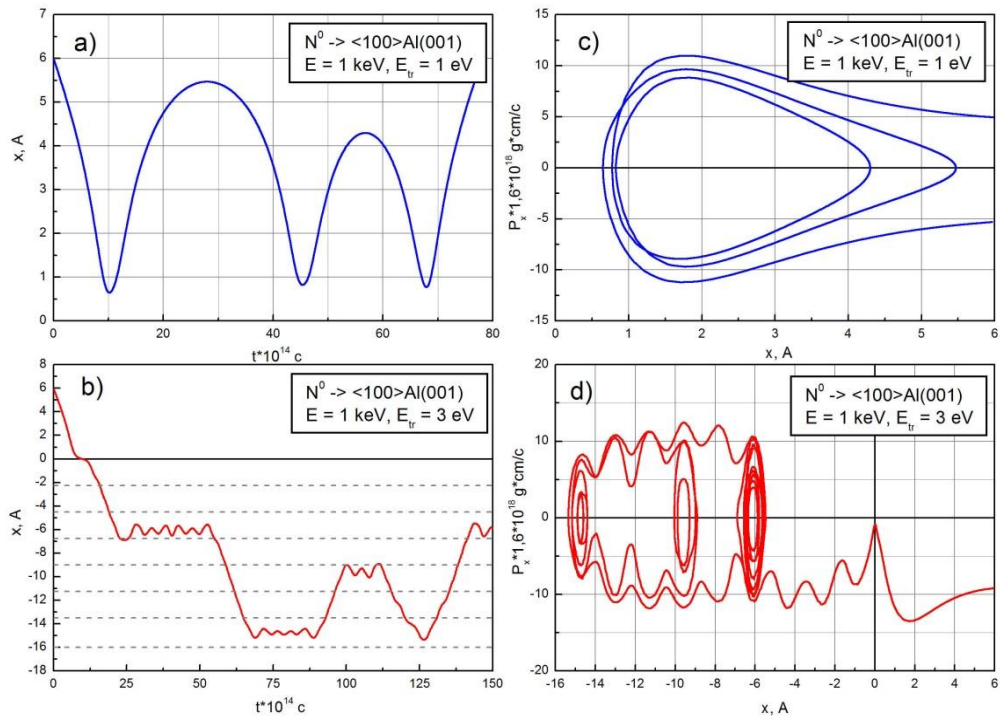

Figure 7

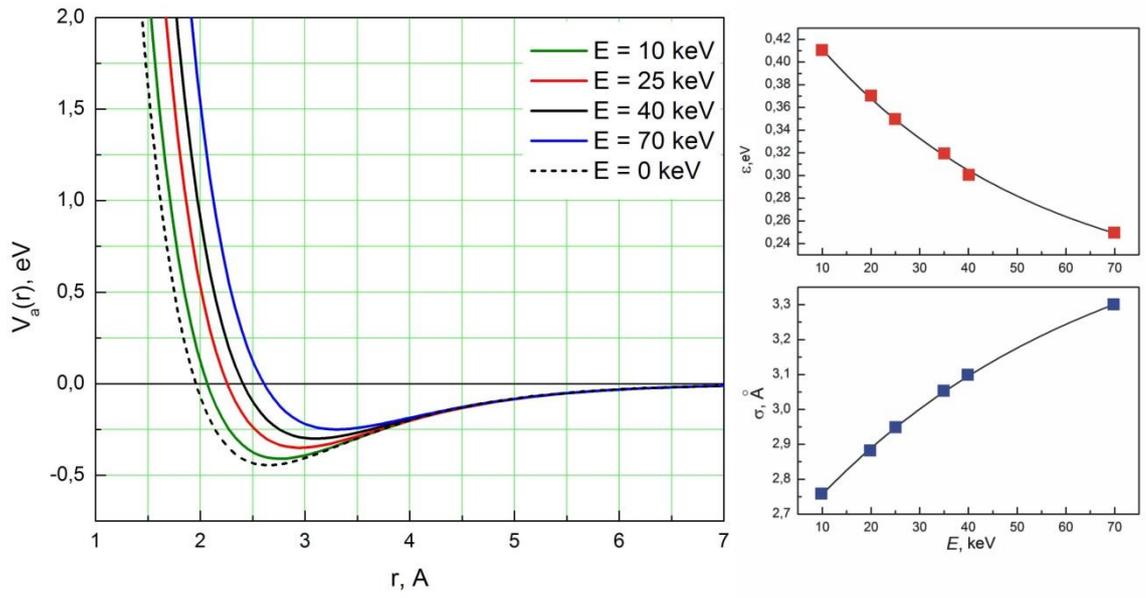

Figure 8

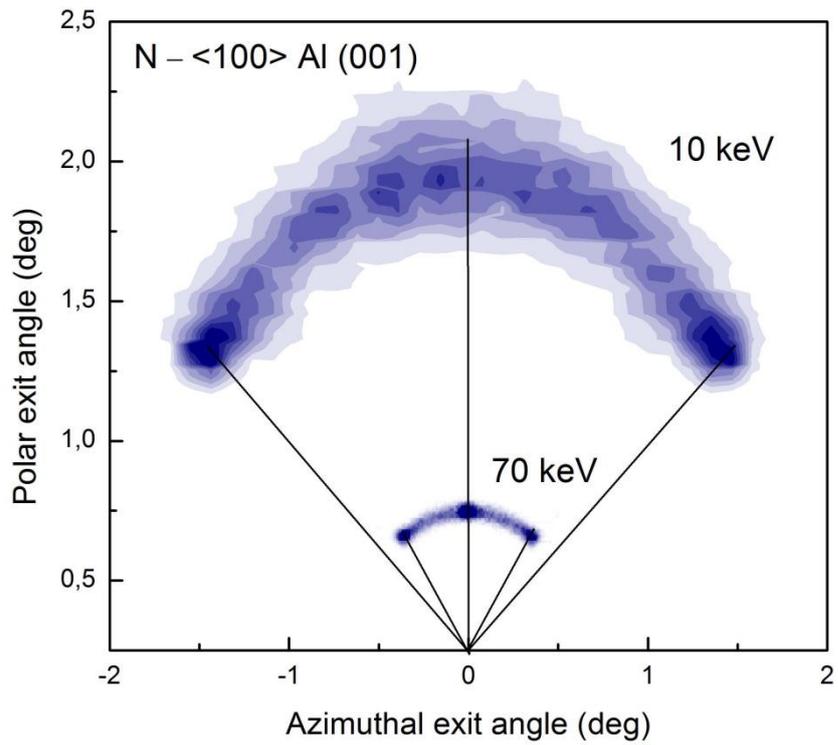

Figure 9

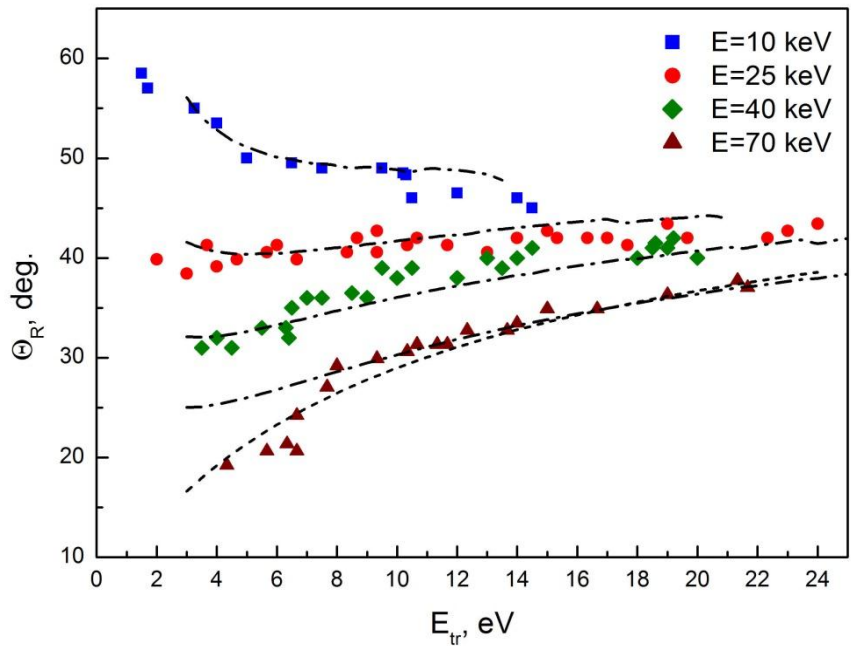

Figure 10

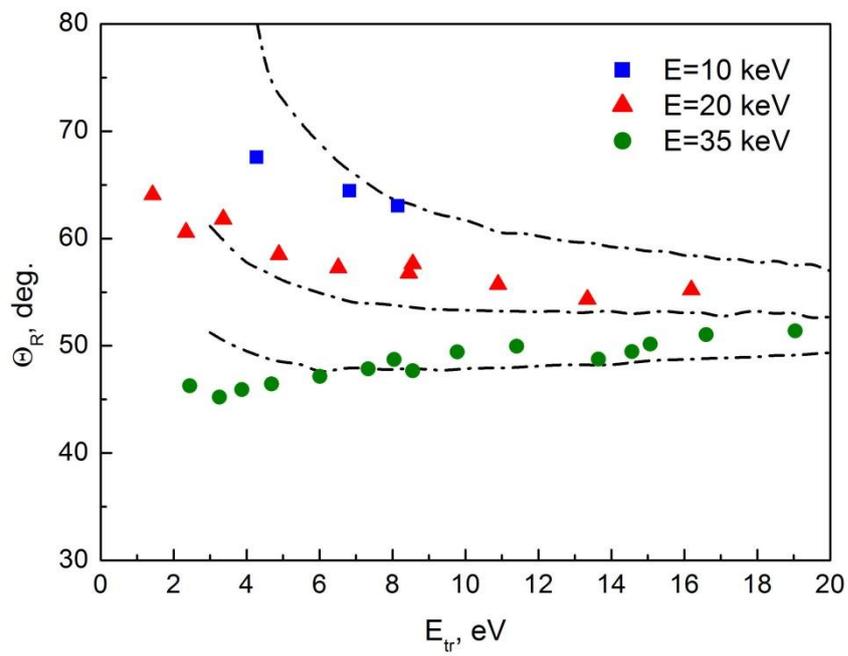

Figure 11